\documentstyle [epsf,epsfig,12pt]{article}  
\begin{document}

\vspace*{-1.8cm}
\hspace*{13cm}{\bf LAL 99-26}\\
\vspace*{-0.5cm}
\hspace*{13.9cm}{May 1999}
\vskip 6.5 cm

\begin{center}
{\LARGE\bf Precise measurement of Higgs decay rate \\
\vskip 2mm
into $W W^*$ at future e$^+$e$^-$ Linear Colliders \\
\vskip 4mm
and theoretical consequences }
\end{center}

\vskip 3 cm
\begin{center}
{\Large\bf G. Borisov, F. Richard }
\end{center}

\vspace*{0.5cm}

\begin{center}
{\large\bf Laboratoire de l'Acc\'el\'erateur Lin\'eaire}\\ 

{IN2P3-CNRS et Universit\'e de Paris-Sud, BP 34, F-91898 Orsay Cedex}
\end{center}

\vskip 1 cm
\begin{center}
{\bf\Large }
\end{center}

\vspace*{-1cm}

\begin{center}
{\bf\large }
\end{center}

\vskip 0.5 cm
\begin{center}
{\it Work presented at the International Workshop on Linear Colliders}\\
{\it Sitges (Barcelona), April 28 - May 5, 1999}
\end{center}
\newpage

\pagestyle{plain}

\section{ Introduction}

 Assuming  a SM or MSSM scenario\cite{review}, one expects a light Higgs 
boson which could be 
studied in great detail with a LC operating at $\sqrt{s}>m_h+m_Z$. In the TESLA
scenario, with 500 fb$^{-1}$ accumulated at $\sqrt{s}=$350GeV, about
10$^5$ hZ events could be produced through the Higgstrahlung process. \par
 At a future LC with a $\sim$ 1 cm beam-pipe radius and
a thin Si detector there will be 
excellent separation between the various flavours\cite{GB}. With the high
statistics available it will thus become possible to measure the various 
branching
ratios with a few $\%$ error. Typically one expects 8 \% precision on 
$BR(h\rightarrow\bar{c}c)$, 6 \%  on $BR(h\rightarrow gg)$ and $\sim 1 \%$  
on $\sigma(hZ) \times BR(h\rightarrow\bar{b}b)$. Furthermore, 
if m$_h>$100 GeV, one will be able to
access to BR(h$\rightarrow WW^*)$ 
and, as explained in section 3, this measurement can give access 
to the Higgs total decay width and
therefore to all partial widths. In particular one can precisely measure 
$\Gamma(h\rightarrow\bar{b}b$) and 
$\Gamma(h\rightarrow\tau^+\tau^-$) which have a high sensitivity 
to MSSM effects\cite{gunion} 
and therefore allow an essential test of
the Higgs sector. If m$_A<$ 1 TeV, it becomes possible to measure a 
significant deviation and, within MSSM, to give an
indirect estimate of m$_A$ thus extending the effective sensitivity above 
the discovery domain of LHC. \par
In the following section we describe a detailed analysis on the 
measurement of  BR(h$\rightarrow WW^*)$.

\section {Measurement of the branching rate $h \rightarrow W W^*$}

In this paper we study the possibility of selecting the decay
$h \rightarrow W W^{*}$ for the mass of $h$ in the range
110 - 140 GeV/$c^2$ with a linear $e^+ e^-$
collider at $\sqrt{s}$=~350~GeV and with an integrated luminosity 
of 500 fb$^{-1}$. The possibility to select this decay mode
for a heavier Higgs boson is discussed elsewhere\cite{w1,w2}.

Both signal and background processes were generated 
using PYTHIA \cite{pythia} version 5.722 with the initial state radiation 
switched on. For the moment we did not take into account the beamstrahlung 
which gives additional $\sim$3\% spread of the centre-of-mass energy and some 
tails. To simulate the detector response, we suppose the following 
parameters of the detector. The charged and neutral particles are registered 
if their momentum is more than 100 MeV/c and the polar angle of their 
direction is $|cos\theta|<0.99$. The efficiency of particle reconstruction 
is 99\%. 
The transverse momentum resolution for the charged particles is: 
$\delta p_t/p_t^2 = 7.\times 10^{-5}(GeV/c)^{-1}$. The energy resolution
for the photons is $\delta E/E = 0.10/\sqrt{E} \oplus 0.01$
and for the neutral hadrons is $\delta E/E = 0.50/\sqrt{E} \oplus 0.04$ 
(E is in GeV).

We consider the reconstruction of 
$h \rightarrow W W^{*}$ in the process $e^+ e^- \rightarrow h Z$
with $W W^{*} \rightarrow l \nu q \bar{q}$ ($l = \mu, e$), 
$Z \rightarrow q \bar{q}$. This final state covers 20.4\%
of all decays of $(h \rightarrow W W^{*}) Z^0$.
The measurement of this $h$ decay in the other final states is also
possible. The main sources of the background are the production of $W^+W^-$,
$Z^0Z^0$ and $t \bar{t}$. The dominating $h$ decay,
$h \rightarrow b \bar{b}$, also gives some contribution
to the background. We find that the contribution of 
$e^+e^- \rightarrow q\bar{q}$ when $q \neq t$ as well as that of 
other processes ($W e \nu$, $Z e e$ etc.) can be reduced to the
negligible level by the topological cuts.

The selection procedure starts by the selection of the lepton
(electron or muon) with an energy above 10 GeV. The remaining particles
are clustered into jets using the JADE algorithm with 
$y_{min} = 0.01$\cite{pythia}.
The events with the number of jets less than 3
are rejected. The transverse momentum of the lepton with respect to 
the nearest jet is required to be greater than 8 GeV/c. The total 
visible mass of all particles excluding the lepton should be in the range:
$130 < M_{vis} < \sqrt{s}-40$ GeV/$c^2$ and the mass of the system
``lepton-missing momentum'' should exceed 10 GeV/$c^2$. 

Some cuts are constructed to reduce the specific types 
of the background. To suppress the background from 
$Z Z^{(*)} \rightarrow l^+l^- q \bar{q}$
the event is removed if the mass of given lepton candidate with any other
lepton is within the $Z^0$ mass ($|M_{ll}-M_Z|<15$ GeV/$c^2$) or less
than 15 GeV/$c^2$. Additionally we reject event if one of jets 
contains only one charged particle or have the mass less than 2 GeV/$c^2$.

The variable $\cos\theta_{vis}\times Q_l$,
where $\theta_{vis}$ is defined as the polar angle of the direction
of the visible momentum (excluding the lepton) and $Q_l$ is the charge
of the lepton, is required to be: 
$-0.95 < \cos\theta_{vis}\times Q_l < 0.90$. This cut reduces the 
background of $W^+W^-$, which is produced in the forward direction, 
and $Z Z^{(*)}$, which is produced both in the forward and 
backward directions. 

\begin{figure}[tbh]
\vspace*{-1.0cm}      
\begin{center} 
  \epsfxsize=16cm
  \epsffile{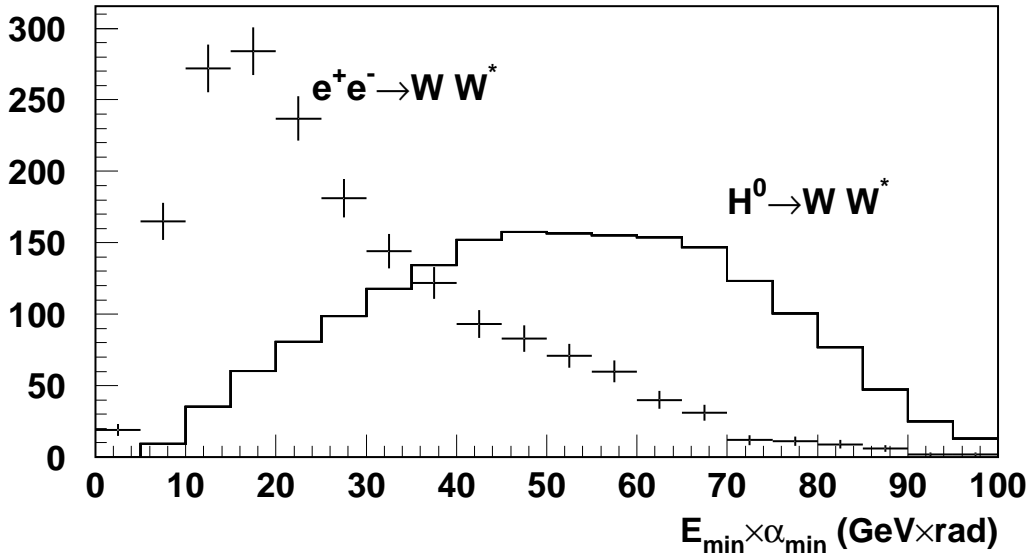}
    \caption{ Distributions of $E_{min} \times \alpha_{min}$ for the
signal (histogram) and for the $W W$ background (points) with the 
arbitrary normalisations.}
    \label{fig:fig1}
\end{center}
\end{figure}

The channel $e^+e^- \rightarrow W^+W^- \rightarrow l \nu q \bar{q}$ has
initially 2 partons so that events with 3 or more jets can arise 
only from gluon emission. To suppress such events the variable sensitive
to the gluon emission can be used. We use the variable
$E_{min} \times \alpha_{min}$ where $E_{min}$ is the minimal jet energy
and $\alpha_{min}$ is the minimal angle between any two jets.
This variable is widely used to suppress events with gluon
emission in LEP experiments.  Its distribution for
$W^+W^-$ and $h Z$ events is shown in Fig.\ref{fig:fig1}. 
Certainly more sophisticated variables from the arsenal of the 
methods developed at LEP can give even better suppression
of this type of background. We reject events if 
$E_{min} \times \alpha_{min} < 45$ (GeV$\times$rad) for events
with 3 jets and $E_{min} \times \alpha_{min} < 20$ (GeV$\times$rad)
for events with 4 and more jets. This variation for the different number
of jets is explained by the fact that the remaining $W^+W^-$ events are 
mainly 3-jet like, while $h Z$ events are more 4-jet like.

\begin{figure}[tbh]
\begin{center} 
  \epsfxsize=16cm
  \epsffile{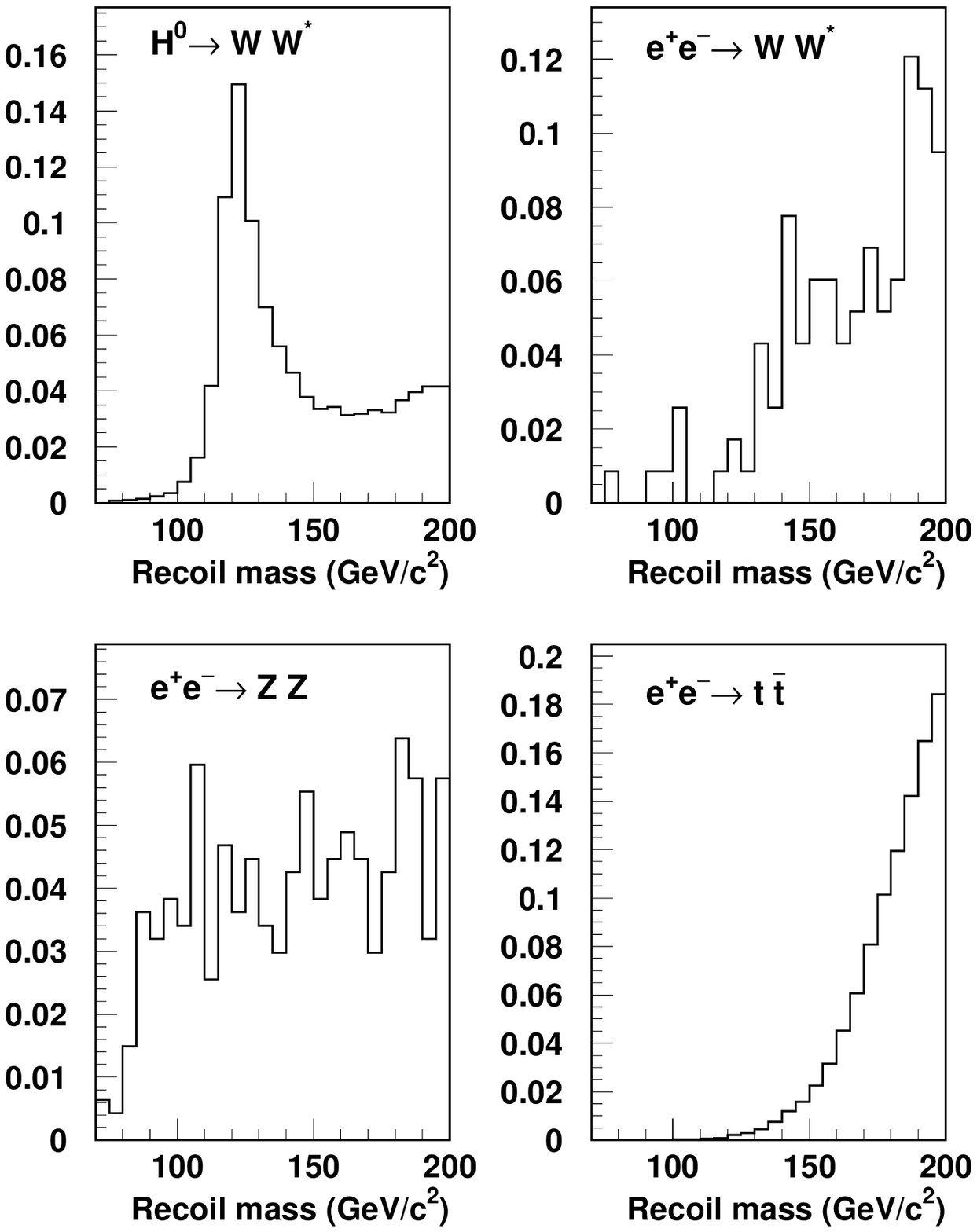}
    \caption{ Distributions of the recoil mass for the signal
with $M_H$= 120 GeV/$c^2$ and for the main types of background.
The normalisation is arbitrary.}
    \label{fig:fig2}
\end{center}
\end{figure}

Finally all particles in the event, excluding the lepton, are forced to 
4 jets and for each possible pairing of jets the mass of the pair
of jets and the recoil mass were computed. The distribution
of the recoil mass when the mass of the pair is within 10 GeV/$c^2$
of the nominal $Z^0$ mass is shown in Fig.\ref{fig:fig2} for the signal 
with $M_H=120$ GeV/$c^2$ and for 3 dominant types of background. 
The normalisation in each case is arbitrary. We select the event if 
the mass of the pair $|M_{jj}-M_Z| < 10$ GeV/$c^2$ and 
the recoil mass $|M_{rec}-M_H-5.0| < 15$ GeV/$c^2$.
This cut gives very strong suppression of $W^+W^-$ (20~times) 
and $t \bar{t}$ ($>200$ times) background. The impact on the signal
is also strong and more than 50\% of the signal events are 
removed by this condition. This effect is explained by undetected ISR
and by errors 
in the jet clustering and in the measurement of the energy flow. 
Some optimisation of the analysis is possible at this stage, however
this cut should be kept in some form to reduce the background of $t \bar{t}$
to a reasonable level. Another alternative is to perform this
measurement below the $t \bar{t}$ threshold.
The remaining number of events for $\int L dt$ = 500 fb$^{-1}$ for
the different types of background and for the signal 
is given in table \ref{tab:tab1}.

\begin{figure}[tbh]
\begin{center} 
  \epsfxsize=16cm
  \epsffile{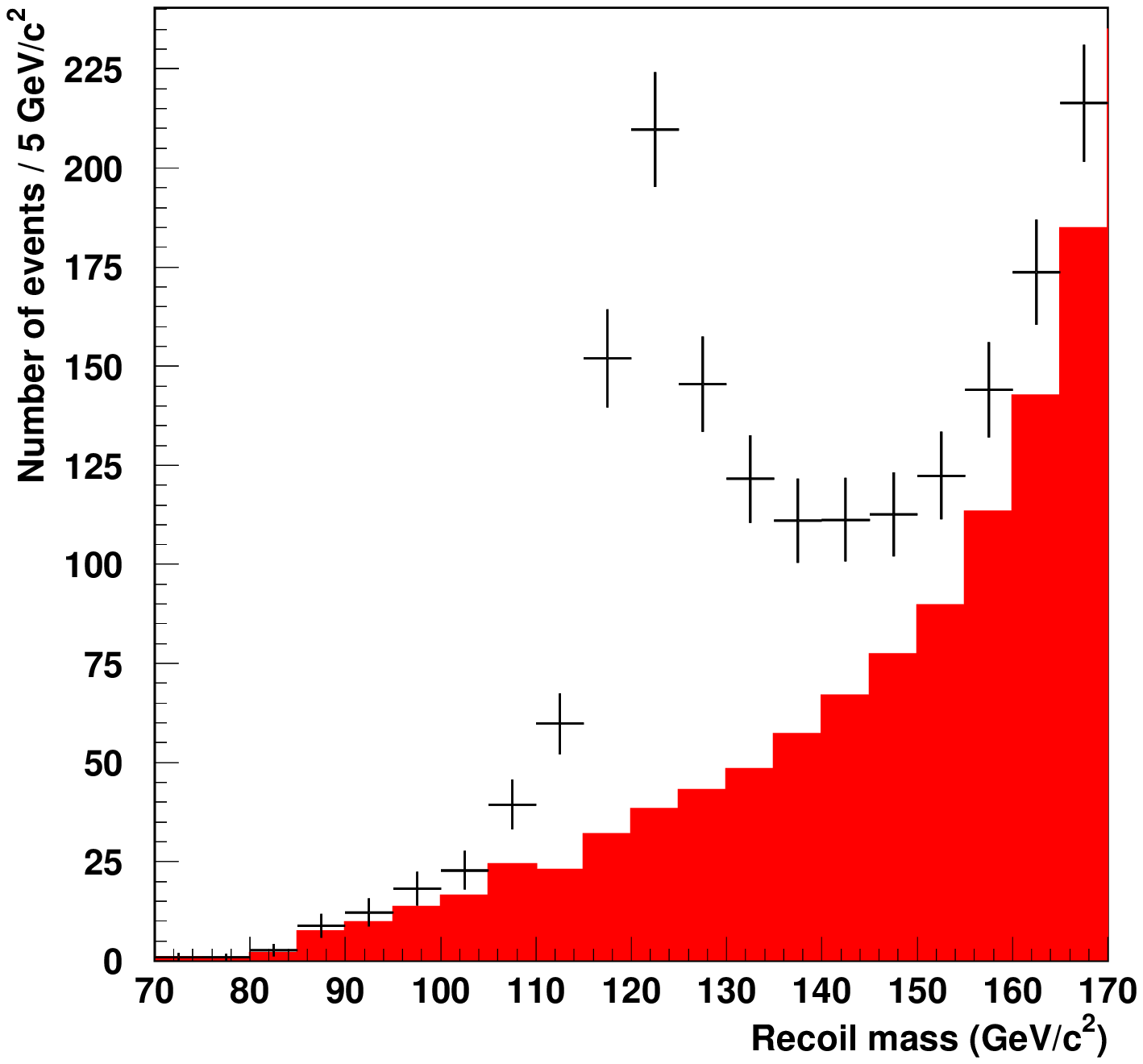}
    \caption{ Expected distributions of the mass recoiling to 
any pair of jets with $|M_{jj}-M_Z| < 10$ GeV/$c^2$. The distribution
is normalised to $\int L dt$ = 500 fb$^{-1}$. The filled histogram
shows the mass distribution for the background. The signal 
$h \rightarrow W W^*$ is generated with $M_H$= 120 GeV/$c^2$.}
    \label{fig:fig3}
\end{center}
\end{figure}

It should be noted that $\sim$100\% of the remaining 
$t \bar{t}$, 85\% of $Z Z$ and 90\% of $h \rightarrow X \neq W W^*$ 
background contains jets with B-hadrons while for the decay
$h \rightarrow W W^*$ B-hadrons can only be produced in the
decay $Z^0 \rightarrow b \bar{b}$ with 
$BR(Z^0 \rightarrow b \bar{b})= 0.154$. This kind of background can
therefore be effectively 
suppressed by applying the anti-b tagging. The precise vertex detector
and the efficient b-tagging algorithms developed for experiments
at LEP and SLC can provide background suppression 
by more than 20 times while keeping the efficiency
for the signal at the 90-95\% level. 
In this study we apply a ``soft'' anti-b tagging which removes
the event if a B-hadron is found in the jets not included into the ``Z$^0$-like''
pair. We suppose that the efficiency of anti-b tagging is 5\% for
an event with 2 B-hadrons and 95\% for the event without B-hadron.
The number of events remaining after this selection is given in
table \ref{tab:tab2} and figure \ref{fig:fig3} shows the 
distribution of the recoil mass for $M_H$=120 GeV/$c^2$.

The only important 
background which remains after above selections is the production 
of $W W^*$ pairs when one of W is off-shell.
There is an interesting possibility to additionally reduce the 
$W W^*$ background using a polarised $e^-$ beam. For the right-handed 
incoming electron with polarisation $P_{e^-}$ the cros-section for
$W W$ production is suppressed approximately by a factor 
$1-P_{e^-}$ while the cros-section of the 
$h$ production is almost unchanged. Therefore an electron polarisation 
$P_{e^-} \sim 0.90$ is sufficient to suppress 
$W W^*$ background and obtain the pure sample of 
$h \rightarrow W W^*$ decays.

There are several ways to improve the precision of the measurement of
$BR(h \rightarrow W W^*)$. We note that the present analysis
covers only 20\% of the $Z^0 H\rightarrow W W^*$ decay modes.
With the level of background reachable with polarisation, one
can probably access to the hadronic W decay modes and/or to 
$Z^0 \rightarrow \nu \bar{\nu}$ etc... Therefore an increase
of the efficiency by a factor $\sim 2$ seems feasible.
We also note that the selection efficiency drops by 50\% due to our
cut on the recoil mass which could be avoided by working below
the $t \bar{t}$ threshold. Thus we can expect that the statistical
errors given in table \ref{tab:tab2} could be reduced by about a factor 2
if necessary. We also expect that the measurement of this decay
mode with $\sim$10\% precision for the Higgs boson with the mass around 
100 GeV/$c^2$ could be possible in an experiment below 
$t \bar{t}$ threshold with a polarised beam.

\begin{table}[tbh]
  \begin{center}
    \begin{tabular}{|c|c|c|c|c|c|c|c|} \hline
$M_H$ & BR(WW) & $H\rightarrow WW$ & $H\rightarrow X\neq WW$ & 
WW & ZZ & $t\bar{t}$ & $(\delta(Br)_{stat}/Br)$(\%) \\ \hline
110   & 0.05 & 152  & 49 & 46  & 71 & 26 & 12.2 \\
120   & 0.14 & 535  & 58 & 116 & 61 & 72 & 5.4  \\
130   & 0.30 & 1280 & 44 & 267 & 55 & 175& 3.3  \\
140   & 0.48 & 2148 & 42 & 371 & 54 & 368& 2.5  \\
\hline
    \end{tabular}
    \caption{Number of selected events for the different processes.
The last column gives the expected statistical precision of the measurement
for $\int L dt$ = 500 fb$^{-1}$. }
    \label{tab:tab1}
  \end{center}
\end{table}

\begin{table}[tbh]
  \begin{center}
    \begin{tabular}{|c|c|c|c|c|c|c|c|} \hline
$M_H$ & BR(WW) & $H\rightarrow WW$ & $H\rightarrow X\neq WW$ & 
WW & ZZ & $t\bar{t}$ & $(\delta(Br)_{stat}/Br)$(\%) \\ \hline
110   & 0.05 & 143  & 3	 & 43  & 18 & 7   & 10.2 \\
120   & 0.14 & 503  & 4	 & 109 & 17 & 20  & 5.1  \\
130   & 0.30 & 1203 & 5	 & 251 & 17 & 45  & 3.2  \\
140   & 0.48 & 2019 & 5	 & 349 & 15 & 99  & 2.5  \\
\hline
    \end{tabular}
    \caption{Number of selected events after applying an anti-b tagging
selection.
The last column gives the expected statistical precision of the measurement
for $\int L dt$ = 500 fb$^{-1}$. }
    \label{tab:tab2}
  \end{center}
\end{table}

\section{ Theoretical Implications}

 The minimal SUSY scenario MSSM, with two Higgs doublets, predicts two 
CP-even Higgs bosons h and H
with a mixing angle $\alpha$, one CP-odd boson A and two charged 
bosons H$^{\pm}$.
MSSM predicts m$_h<$130 GeV, while no clear upper bound is given for 
the rest of the
spectrum. In the mSUGRA and gauge-mediated scenarios one generally 
expects that these
particles will be heavy\cite{erler} and therefore not directly 
observable at the first stage of
a LC operating at $\sqrt{s}<$500 GeV. In the appendix, we recall 
why $\Gamma(h\rightarrow\bar{b}b$)
has a high sensitivity to MSSM provided m$_A<$1 TeV. This quantity 
cannot be directly measured but requires
a combination of at least two measurements. As discussed below, 
a precise measurement is only feasible if
the channel h$\rightarrow WW^*$ is experimentally accessible. This 
is possible
when m$_h>$ 100 GeV, a scenario becoming increasingly probable 
with the LEP2 limits reaching 95 GeV.
Nevertheless, for the sake of comparison, two scenarios will be 
discussed. \par
1/ BR($WW^*$) not accessible \par
 One can measure $\sigma$(hZ) inclusively (using the Z 
leptonic decays) with the precision $\sim$2\%
and $\sigma$(hZ)BR(h$\rightarrow\bar{b}b)$. BR(h$\rightarrow\bar{b}b)$ 
has some
sensitivity to m$_A$, but this sensitivity is reduced with respect 
to $\Gamma(h\rightarrow\bar{b}b$) since
the total width itself is dominated by the same contribution and since 
the extra contributions to the total width
coming from $\Gamma(h\rightarrow\bar{c}c$) and $\Gamma(h\rightarrow gg$) 
are not accurately computable. \par
 BR(h$\rightarrow\bar{b}b)$/BR(h$\rightarrow\bar{c}c)$ is also accurately 
measurable and, as shown in the appendix,
has the same sensitivity to MSSM as $\Gamma(h\rightarrow\bar{b}b$). 
It turns out however that 
$\Gamma(h\rightarrow\bar{c}c)\sim m^2_c(m_h)$ is poorly known at 
the theoretical level. This translates into an uncertainty on
$\Gamma(h\rightarrow\bar{c}c$) of at least 10$\%$ which reduces 
the sensitivity on m$_A$ to $\sim$ 500 GeV. \par
One can alternatively access to the total decay width $\Gamma_T$ and 
then derive
$\Gamma(h\rightarrow\bar{b}b$)=BR(h$\rightarrow\bar{b}b)\Gamma_T$. To 
do this, one uses $\Gamma(h\rightarrow\gamma\gamma$)
measured with a $\gamma-\gamma$ collider and BR(h$\rightarrow\gamma\gamma$) 
measured in e$^+$e$^-$.  Recent\cite{brient}
estimates however predict a statistical accuracy on 
BR(h$\rightarrow\gamma\gamma$) not better than 15 $\%$. \par
In conclusion this scenario, even with the TESLA luminosity, does not 
allow to reach an accuracy better than 10 $\%$ on
$\Gamma(h\rightarrow\bar{b}b$). \par
2/ BR($WW^*$) accessible \par 
 In this scenario one can directly measure 
$\Gamma$(h$\rightarrow\bar{b}b)$/$\Gamma$(h$\rightarrow WW^*)$ which, 
as shown in the appendix, 
has the same sensitivity to m$_A$ as $\Gamma(h\rightarrow\bar{b}b$). This 
simple minded statement assumes MSSM in which the
Higgs boson has a SM-like coupling to vector bosons. This assumption 
can however be tested in two complementary ways.
One can test the h-Z-Z coupling via the measurement of the 
Higgstrahlung cross-section $\sigma(Z^*\rightarrow hZ)$ and assume
universality with the h-W-W coupling. The latter assumption can be checked 
by measuring the fusion process 
e$^+$e$^-$ $\rightarrow\nu\bar{\nu}W^*W^*\rightarrow$h which allows 
to accurately verify universality.  \par
   After performing these checks, if no deviation is observed 
on the h-W-W coupling, one can safely compare
   $\Gamma$(h$\rightarrow\bar{b}b)$/$\Gamma$(h$\rightarrow WW^*)$ to 
the MSSM predictions and derive a limit on m$_A$. Table \ref{tab:tab3}
indicates the type of accuracy which can be reached for 4 Higgs masses. 
As noted previous section a dedicated measurement performed
below the top threshold could allow an improved statistical 
accuracy by about a factor 2. 
\begin{table}[t]
\centering
\vskip 0.5 truecm
\begin{tabular}{|c|c|c|c|c|c|}
\hline m$_h$ & BR($WW*)$ & $b\bar{b}/WW^*$   & $\Gamma_T$ &
 95 \% C.L. m$_A$  \\
 GeV         &  \%       &   \%                                 &
     \%     & GeV      \\   
\hline 110       & 5  &  10  &   10   &  550   \\
\hline 120      & 14     & 5 &    5  &  750  \\
\hline 130       & 30 &  3        & 3.6  &   1000 \\
\hline 140       & 48 &  2.5         & 3.2  &   1100 \\
\hline
\end{tabular}
\caption{Typical precision for Higgs branching ratios with 500 fb$^{-1}$ 
at 350 GeV}
\label{tab:tab3}
\end{table}

Figure \ref{fig:fig4} indicates the corresponding sensitivity which
can be reached on m$_A$. In a favourable case, say m$_h$=120 GeV, the 
statistical accuracy
is sufficient up to m$_A\sim$ 1 TeV. \par
 This sensitivity can be compared to the LHC discovery reach. 
Direct observation of a heavy A
is only possible at low $tan\beta$ (region excluded if no Higgs is 
found with mass below 100 GeV)
or very high $tan\beta$ through the decay of A into $\tau^+\tau^-$. 
This leaves a large interval which 
a precise measurement of $\Gamma$(h$\rightarrow\bar{b}b)$ would 
allow to cover. \par
 At this stage one should take into account the various systematical
errors. On the theoretical side, the most obvious effect comes from 
the uncertainty on the b quark mass. At present the error
is $> 1\%$ and therefore this effect can be relevant. A possible 
way out is to use the measurement of
$\Gamma$(h$\rightarrow\tau^+\tau^-)$ for which the statistical error 
will probably be worse. On the experimental side, one should
note that at LEP2 the typical efficiency on the $\tau^+\tau^-$ channel 
is about 20 $\%$ but this comes from the limited
vertex accuracy of LEP2 detectors which are unable to detect vertex 
offsets from $\tau$ particles while this would not be the case 
at future LC. \par  
 Another source of uncertainty comes from theoretical inputs, like 
for instance the radiative corrections. As discussed in
the appendix, it seems that the dependence on unknown parameters is weak 
provided
that $tan\beta$ is above 2 (or equivalently that m$_h$ is 
above 100 GeV). This statement is probably
too naive and deserves further investigations. \par
   Finally, if no deviation is observed on the h-W-W to coupling, one 
can also access to the total Higgs decay width using
    $\Gamma_T=\Gamma(WW^*)/BR(WW^*)$, where $\Gamma(WW^*)$  is a 
computable quantity and $BR(WW^*)$ is obtained from the measurements
of $\sigma(hZ)\times BR(WW^*)$ and $\sigma(hZ)$. The total Higgs decay 
width could therefore be measured with a much higher precision than by using 
the $\gamma-\gamma$ channel. The numbers given in table 3 assume the 
error on $\sigma(hZ)$ of 2\%.
\par   

\section { Conclusion}

A precise measurement of BR(h$\rightarrow WW^*)$, possible if 
m$_h>$100 GeV, allows to measure
precisely the Higgs total decay width and therefore access to 
$\Gamma(h\rightarrow\bar{b}b$). This
extends considerably our ability to discriminate between the SM and 
MSSM scenario and therefore illustrates the potential
of precise measurements of Higgs branching ratios.
\vskip 2. cm
{\Large \bf Appendix}
\vskip 1.0 cm
 The following calculations give a simplified treatment only meant 
to understand the dependence of various observables in
terms of the MSSM parameters. \par
 One has $\Gamma(h\rightarrow\bar{b}b) = 
\Gamma_{SM}(h\rightarrow\bar{b}b)sin^2\alpha/cos^2\beta$ where $\alpha$, 
varying
between -$\pi$/2 and 0, defines the mixing between the two CP-even 
Higgs bosons and $\beta$, varying between $\pi$/4 and $\pi$/2,
is defined from the ratio of the vacuum expectations of the two doublets. \par
 This two angles are related through the formula :
  $$tan2\alpha=tan2\beta\frac{m^2_A+m^2_Z}{m^2_A-m^2_Z+\epsilon/cos2\beta} $$
with :
  $$ \epsilon=\frac{3G_F}{\sqrt{2}\pi^2}\frac{m^4_t}{sin^2\beta} 
Log(m_{\tilde{t}_1}m_{\tilde{t}_2}/m^2_t) $$
where $m_{\tilde{t}_{1,2}}$ are the two stop masses. \par
  This is an approximate formula, ignoring mixing effects, but good enough 
to get a first guess of the relevant effect. \par
 When m$_A>>m_Z$, this formula shows that 
$\beta-\alpha\rightarrow\pi/2-\eta$ where $\eta$ is small. One can easily
derive that :
 $$ \eta=\frac{m^2_Z|cos2\beta|+\epsilon/2} 
{m^2_A-\epsilon/|cos2\beta|}sin2\beta $$
 One has :
 $$ \frac{sin^2\alpha}{cos^2\beta}\rightarrow 1-2\eta tan\beta $$
 Similarly one can estimate the change on the hZ cross-section :
 $$ sin^2(\beta-\alpha)\rightarrow 1-\eta^2$$
 From this one concludes that there is a linear dependence on $\eta$ for 
the width while the cross-section
has a quadratic dependence
therefore vanishing quickly for large m$_A$. \par
\begin{figure}
\epsfysize15cm
\epsfxsize15cm
\epsffile{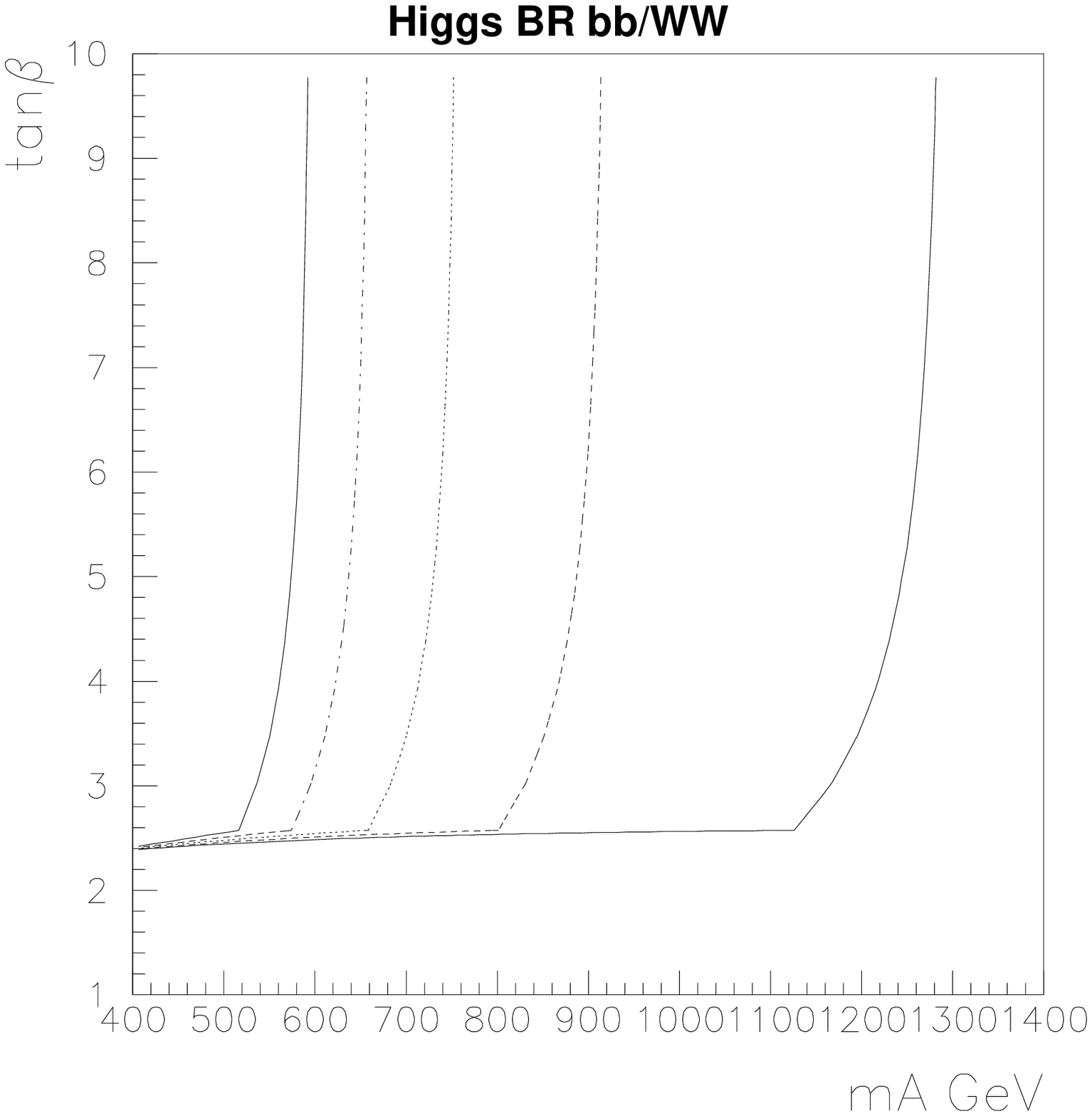}
\caption{ MSSM effect on BR(h$\rightarrow b\bar{b}$)/BR(h$\rightarrow WW^*)$
for m$_h$=110 GeV.
The 5 curves correspond to 1.03,1.06,1.09,1.12 and 1.15 correction factors.}
\label{fig:fig4}
\end{figure}

Figure \ref{fig:fig4} shows the behaviour of 
$\Gamma(h\rightarrow\bar{b}b$)/$\Gamma(h\rightarrow WW^*$). One observes the
same behaviour as in \cite{gunion}. The region at low $tan\beta$ is cut 
away by requesting that the MSSM parameters are 
compatible with a Higgs mass of 110 GeV. \par
Note that $\Gamma(h\rightarrow\tau^+\tau^-$) has the same correction 
factor than $\Gamma(h\rightarrow\bar{b}b$).\par
 For what concerns $\Gamma(h\rightarrow\bar{c}c$), the correction factor :
  $$ \frac{cos^2\alpha}{sin^2\beta}\rightarrow 1+2\eta/tan\beta $$
  This expression allows to understand why the effect coming from 
$\Gamma(h\rightarrow\bar{c}c$) is reduced : since $\eta$=0
for $tan\beta$=1, this effect is only relevant for larger values of 
$tan\beta$ but then it is damped with respect to the 
corresponding effect on $\Gamma(h\rightarrow\bar{b}b$).  \par
\begin{figure}
\epsfysize15cm
\epsfxsize15cm
\epsffile{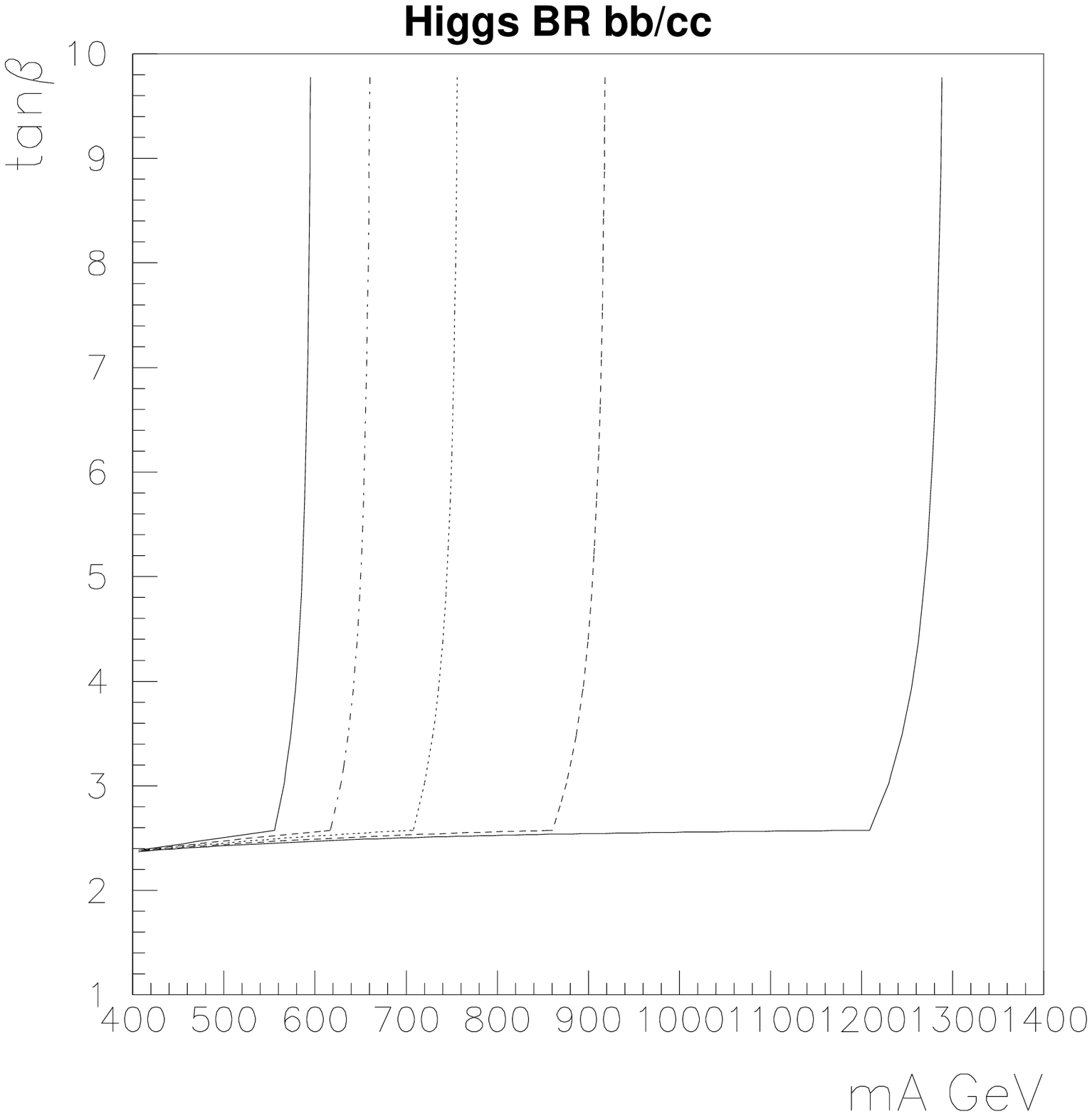}
\caption{ MSSM effect on BR(h$\rightarrow b\bar{b}$)/BR(h$\rightarrow c\bar{c})$
for m$_h$=110 GeV.
The 5 curves correspond to 1.03,1.06,1.09,1.12 and 1.15 correction factors.}
\label{fig:fig5}
\end{figure}

  Figure \ref{fig:fig5} shows the behaviour of 
$\Gamma(h\rightarrow\bar{b}b$)/$\Gamma(h\rightarrow \bar{c}c$).  \par  
 For large m$_A$, the Higgs mass is given by :
 $$ m^2_h=m^2_Zcos^22\beta+\epsilon sin^2\beta $$
 One can therefore express the term $\eta tan\beta$, which defines the 
correction effect for $\Gamma(h\rightarrow\bar{b}b$), 
 in terms of $m^2_h$ :
  $$ \eta tan\beta=-\frac{m^2_Z|cos2\beta|+m^2_h}
{m^2_A-\epsilon/|cos2\beta|} $$
  If we assume that $tan\beta>2$ and that m$_A$ is large, one can drop 
the $cos2\beta$ dependence with no significant
loss of precision:
   $$ \eta tan\beta=-\frac{m^2_Z+m^2_h}{m^2_A} $$
  meaning that the correction factor is essentially independent 
of $tan\beta$, as can be inferred by the curves of figure \ref{fig:fig4}.
  Present LEP2 limits, within some assumptions, tend to exclude the 
low $tan\beta$ region and it seems therefore that this
  approximation will apply.


\end{document}